# Magnetic Energy Injection in GRB 080913

Liu Xue-Wen[†], Wu Xue-Feng & Lu Tan

Purple Mountain Observatory, Chinese Academy of Sciences, Nanjing, 210008, China

GRB 080913, with a spectroscopically determined redshift of z=6.7, was the record holder of the remotest stellar object before the discovery of the recent gamma-ray burst GRB 090423, whose redshift is about 8.2. The gradually accumulated high redshift GRB sample has shed light on the origin and physics of GRBs during the cosmic re-ionization epoch. We here present a detailed numerical fit to the multi-wavelength data of the optical afterglow of GRB 080913 and then constrain its circum-burst environment and the other model parameters. We conclude that the late optical/X-ray plateau at about one day since the burst is due to the Poynting-flux dominated injection from the central engine which is very likely a massive spinning black hole with super strong magnetic fields.

gamma-rays: bursts, outflow, black hole

## 1  Introduction

GRB 080913 was discovered by the Swift satellite on 13 Sep 2008, at $T_0$ = 06:46:54 UT with a duration of $T_{90}$ = 8 ± 1 s. The total fluence in the 15-150 keV band is $(5.6 \pm 0.6) \times 10^{-7}\,\mathrm{erg\,cm^{-2}}$. Fit to the combined data of Swift/BAT (15-150 keV) and Konus-Wind (20-1300 keV) by a Band function ($\beta_{high}$ = -1.5 fixed) yields $\beta_{low}$ = -0.28 $^{+0.75}_{-0.53}$ and $E_{peak}$ = 121 $^{+232}_{-39}$ keV, where $\beta_{low}$ and $\beta_{high}$ are the low- and high-energy spectral indices, and $E_{peak}$ is the peak energy [1]. The redshift of z=6.7 was determined by the GROND multi-color photometry and VLT spectra [1,2]. For a standard cosmology model with $H_0$ = 71 km s$^{-1}$ Mpc$^{-1}$, $\Omega_M$ = 0.27 and $\Omega_\Lambda$ = 0.73, the isotropic energy release $E_{iso}$ is $7 \times 10^{52}$ erg (1 keV - 10 MeV, in the GRB progenitor local frame).

Swift/XRT started to obtain X-ray data at 94 s after the BAT trigger time $T_0$. The X-ray spectrum from ~108 s to ~1920 s is well fit by an absorbed power law with photon index Γ=1.66 ± 0.14. However, the early afterglow is likely to be contaminated by flare emission [1]. In the time interval between $T_0$+100 s and $T_0$+10$^5$ s, the X-ray afterglow exhibited a long decay with temporal index $\alpha \sim 1.2$ ($F_\nu \propto \nu^{-\beta} t^{-\alpha}$) superimposed by a likely small flare at ~400 s and a substantial flare at 2 ks with a factor of ~5 increase in flux. After that there appeared an X-ray bump. The late spectrum achieved by XMM-Newton at 380 ks was best fit with photon index Γ=2.0 ± 0.2.

The rapid localization of GRB 080913 by Swift enabled instant responses of the ground-based telescopes such as the REM telescope and the GROND mounted at the 2.2 m MPI/ESO telescope at La Silla [1,3]. The GROND performed simultaneous seven-channel observations about 6 minutes after $T_0$ and obtained up-limits at g'r'i' bands and achieved rich data at z'JHK bands [1]. The optical/NIR light curve can be described by a power-law decay with index ~1.0 up to $T_0$+10$^4$ s and a subsequent prolonged plateau which was suggested by a measurement at ~180 ks with almost the same brightness at $T_0$+10$^4$ s. Late observations by GROND, VLT, NTT, Gemini-N and Subaru showed variability similar to X-rays at nearly the same time and indicated that the best-fit power law index of the near infrared (NIR) spectral energy distribution (SED) is $\beta_{opt} = 1.12 \pm 0.16$.

The overall optical/NIR light curve can be fit by a double broken power law of $\alpha_{opt,1} = 1.03 \pm 0.02$ for the initial decay phase ($t < t_{b,1} \sim 10^4 s$), and $\alpha_{opt,2} = -0.19_{-0.16}^{+0.15}$ for the plateau phase ($t_{b,1} < t < t_{b,2} = 1.16_{-0.17}^{+0.19} \times 10^5 s$), and finally $\alpha_{opt,3} = 0.92_{-0.08}^{+0.09}$ for the post-plateau phase. Although the shape of the X-ray light curve does not match the optical/NIR quite well, using the data of the optical and X-ray for a joint fit does not change the above values significantly [1]. According to the standard afterglow model, the closure relation $\alpha_{opt,1} - 1.5\beta_{opt} \sim -0.5$ favors a constant density environment and the optical band above the cooling frequency $\nu_c$ and the typical synchrotron frequency $\nu_m$. Thus the electron energy distribution index $p \sim 2.2$ can be inferred from the theoretical relations $p = (4\alpha_{opt,1} + 2)/3$ and/or $p = 2\beta_{opt}$.

The similar light curves and SEDs indicate that the optical and the X-ray emission have the same origin. The nearly achromatic beginning and ending of the plateau phase observed in different energy bands suggest that the re-brightening is due to a continuous energy injection into the forward shock. Greiner et al. (2009) used an analytical method to show that the Poynting flux injection is more favorable than the baryonic energy injection. In this paper we use a numerical model to fit the optical/NIR light curves and discuss the implications of our numerical results.

## 2 Numerical Model

The early afterglow ($t < t_{b,1}$) is due to the forward shock when the GRB outflow propagates into the circum-burst medium. A Poynting flux with luminosity $L = L_0(1 + t/T_*)^{-2}$ [4], is assumed to be continuously injected into the forward shock from the central engine (a spinning-down magnetar or black hole). It begins to determine the evolution of the forward shock at $t_{b,1}$, when the injected energy is comparable to the initial kinetic energy. When Poynting-flux energy injection is taken into account, the equation of the generic model for GRB outflow dynamics during the afterglow phase proposed by Huang et al. (2000) [5] can be modified to be [6]

$$\frac{d\gamma}{dm} = -\frac{(\gamma^2 - 1) - \frac{1-\beta}{\beta c^3} \Omega_j L(t - R/c) \frac{dR}{dm}}{M_{ej} + 2(1-\varepsilon)\gamma m + \varepsilon m}, \quad (1)$$

where $\Omega_j = (1 - \cos\theta_j)/2$ is the beaming factor of the GRB jetted outflow, $\theta_j$ is the half-opening angle of the jet, $M_{ej}$ is the total mass of the jet, $m$ is the external mass swept up by the shock, $\varepsilon$ is the radiative efficiency, $R$ is the radius, and $\gamma = 1/\sqrt{1-\beta^2}$ is the Lorentz factor of the forward shock. The characteristic luminosity $L_0$ and timescale $T_*$ of a magnetar and black hole (BH) are [4, 7]

$$L_0 = \begin{cases} 4 \times 10^{47} B_{\perp,14}^2 R_{M,6}^6 P_{0,ms}^{-4} & \text{erg s}^{-1} \text{ (Magnetar)} \\ 3 \times 10^{47} B_{BH,15}^2 \left(\frac{M_{BH}}{3M_\odot}\right)^2 \left(\frac{a}{0.2}\right)^4 & \text{erg s}^{-1} \text{ (BH)} \end{cases} \quad (2)$$

$$T_* = \begin{cases} 0.58 B_{\perp,14}^{-2} I_{45} R_{M,6}^{-6} P_{0,ms}^2 & \text{days (Magnetar)} \\ 1.0 B_{BH,15}^{-2} \left(\frac{M_{BH}}{3M_\odot}\right)^{-1} \left(\frac{a}{0.2}\right)^{-2} & \text{days (BH)} \end{cases}, \quad (3)$$

where $B$ is the magnetic field strength. The magnetar has a radius $R_M$, a momentum of inertia $I$ and a rotation period $P_0$. The rotation parameter of the BH is $a$. We adopt the convention of $Q_x = Q/10^x$ together with the cgs units in this paper. The theoretical temporal index of the plateau phase assuming $L = L_0(1 + t/T_*)^{-2}$ is $(p-2)/2 > 0$ ($p \sim 2.2$) for $t < T_*$, which is inconsistent with the observed value ($\alpha_{opt,2} < 0$). So we assume the following Poynting-flux luminosity (in the rest frame of the central engine)

$$L = \begin{cases} L_0(t/T_*)^q, & t < T_* \\ L_0(t/T_*)^{-2}, & t > T_* \end{cases}, \quad (4)$$

which has the same asymptotic form as that of a spinning-down magnetar or BH after $T_*$. Then we follow the procedure described in Huang et al. (2000) and numerically fit the NIR afterglow of GRB 080913. Our result is shown in Fig. 1, which demonstrates that the afterglow at JHK bands is well fit while the z' band is strongly affected by the Ly-$\alpha$ absorption at high redshift. The parameter values of our best fit are presented in Table 1.

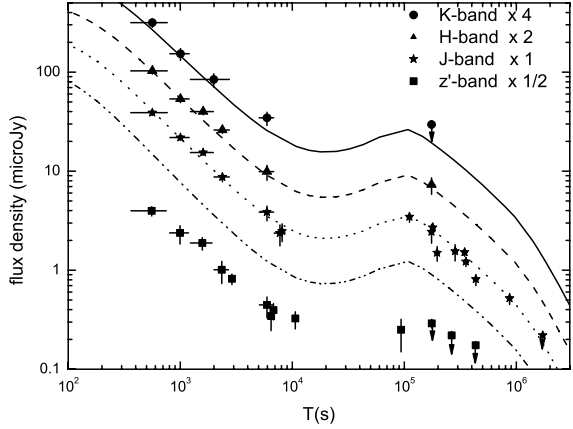

**Fig. 1** Numerical fit to the NIR afterglow of GRB 080913. The data are taken from Greiner et al. (2009).

## 3 Discussions

We have applied the numerical Poynting-flux energy injection model to the NIR afterglow of the high-redshift gamma-ray burst GRB080913. We have showed that in order to explain the plateau of the light curve, the Poynting luminosity is required to be more than $10^{50}$ erg s$^{-1}$ which is impossible even for a millisecond rotating magnetar, but can be accommodated in a fast-rotating massive BH model. Our best fit indicates that GRB 080913 was surrounded by a dense interstellar medium (ISM) with $n = 3000$ cm$^{-3}$, which is consistent with the expectation of ISM at high redshift [8]. Our best numerical fit also shows that the luminosity increases with time as $t^q$ ($q=1.0$) for times earlier than the central engine spin down time scale ($t<T_*$), which is different from the analytical result $q\sim 0.0$ (Greiner et al. 2009). This increase of luminosity may be possibly caused by a considerable amount of the stellar material continually falling back onto the central engine with an increasing mass rate. The total isotropic Poynting energy supplied by the BH is $\sim L_0 T_* \sim 2\times 10^{54}$ ergs which is not unreasonable although quite large.

Although GRB 080913 satisfies the Amati relation, the duration of this burst in the local frame is less than 2 s and its spectral lag is short, both of which indicate that it may be an intrinsically short burst. While according to the flow chart of the criteria for physical classifications proposed by Zhang et al. (2009) [9], GRB080913 is possibly a Type II (collapsar origin) burst. The forthcoming observations are expected to identify the ambiguous situation since the discovery of GRB 090423 has confirmed again the superiority of GRBs as a probe of high-z cosmology.

**Table 1** Best-fit parameter values of GRB 080913

| $E_{k,\mathrm{iso}}$ (ergs) | $\theta_j$ (rad) | $n$ (cm$^{-3}$) | $\gamma_0$ | $p$ |
|---|---|---|---|---|
| $3.7\times 10^{52}$ | 0.22 | 3000 | 3000 | 2.2 |
| $L_0$ (erg s$^{-1}$) | $T_*$ (s) | $\varepsilon_e$ | $\varepsilon_B$ | $q$ |
| $2.4\times 10^{50}$ | $7.8\times 10^3$ | 0.04 | $10^{-5}$ | 1.0 |

**Note:** $E_{k,\mathrm{iso}}$ and $\gamma_0$ are the isotropic kinetic energy and the initial Lorentz factor of the jet which satisfy the relation $E_{k,\mathrm{iso}}\Omega_j = \gamma_0 M_{\mathrm{ej}} c^2$, $\varepsilon_e$ and $\varepsilon_B$ are the energy fractions of electrons and magnetic fields, respectively.

Received; accepted
doi:
†Corresponding author (email: astrolxw@gmail.com)
Supported by the National Basic Research
Propram of China (973 Program, grant 2009CB824800),
the National Natural Science Foundation of China (Grant No. 10473023, 10503012, 10621303, and 10633040), and the Special Foundation for the Authors of National Excellent Doctorial Dissertations of P. R. China by Chinese Academy of Sciences. XFW also thanks the supports of NASA grant NNX08AL40G.